\RequirePackage{fix-cm}
\documentclass[twocolumn,epjc3]{svjour3}
\smartqed  
\RequirePackage{graphicx}
\usepackage{epsfig}
\usepackage{color}
\usepackage{subfigure}
\usepackage{amsmath}
\usepackage{amssymb}
\usepackage{amsmath}
\usepackage{amssymb}
\usepackage{latexsym}
\usepackage{epsfig}
\usepackage{multirow}

\begin{document}

\title{Effect of phantom dark energy on the holographic thermalization}

\author{Xiao-Xiong Zeng \thanksref{addr1}
        \and
  Xin-Yun Hu   \thanksref{addr1}  \and
  Li-Fang Li (corresponding author)\thanksref{e3, addr2}
   }


\thankstext{e3}{e-mail: lilf@itp.ac.cn}
\institute{ School of Science, Chongqing Jiaotong University, Chongqing, 400074,
 China \label{addr1}\and
State Key Laboratory of Space Weather,
Center for Space Science and Applied Research, Chinese Academy of Sciences,
Beijing, 100190, China \label{addr2}}




\maketitle

\begin{abstract}
Gravitational collapse of a  shell of charged dust surrounded by the phantom dark energy is probed by the minimal area surface, which is dual to probe the thermalization in the boundary
quantum field by expectation values of Wilson loop in the framework of the  AdS/CFT correspondence. We investigated mainly the effect of the  phantom dark energy parameter and chemical potential
on the thermalization. The result shows that the smaller the phantom dark energy parameter is, the easier the plasma thermalizes as the chemical potential is fixed, and the larger the chemical potential is, the harder the plasma thermalizes as the dark energy parameter is fixed.  We get the fitting function of the thermalization curve and with it, the thermalization velocity and thermalization acceleration are discussed.

\end{abstract}


\section{Introduction} \label{sec:introduction}
Recent years, more and more theoretical physicists have paid their attention on the applications of the  AdS/CFT correspondence\cite{Maldacena1998}.
It  can not only  check the  effectiveness of this correspondence indirectly, but also provide a method to deal with some problems in strongly coupled system\cite{Sonner2012,li030,Murata2010,Mukhopadhyay2013,Arnab,Shin2012}. One of the strongly coupled systems is the quark gluon plasma which are produced  in heavy ion colliders such as the RHIC and LHC. The properties of the quark gluon plasma have been investigated extensively and we know that it behaves as an ideal fluid with a very small
shear viscosity over entropy density ratio\cite{entropy}. But the process of formation of quark gluon  plasma after a heavy ion
collision, often referred to as thermalization, is not well understood until now.
Recently  some physicists have devoted themself  to addressing this problem from the viewpoint of holography, which is called holographic thermalization. The most prominent property for the thermalization is that it is a non-equilibrium process.
From the AdS/CFT correspondence, we know that the initial state before the thermalization is dual to a pure AdS, and the last state after the thermalization is dual to a stationary black hole. Therefore, to describe the thermalization process, one should construct a dynamical background in the bulk, which can be described as black hole formation or black hole merger.

Now, there have been many models to study this  far-from-equilibrium thermalization behaviors\cite{Garfinkle84,Garfnkle1202,Allais1201,Das343,Steineder,Wu1210,Gao1,Buchel2013,Keranen2012}. A slightly different and simpler model on this topic was put forward
by Balasubramanian et al.\cite{Balasubramanian1,Balasubramanian2}, where the dynamical background was treated as the gravitational collapse of a shell of dust. They claimed that the thermalization process can be probed by the equal-time two-point correlation functions
of local gauge invariant operators, expectation values of Wilson loop operators, and
entanglement entropy, which in the gravity side correspond to minimal lengths, areas, and
volumes in AdS space, respectively. An important conclusion from their model is  that the thermalization time is closer to the
experiment data produced in RHIC and LHC. In addition, they found that the thermalization is a top-down
process, in contrast to
the predictions of bottom-up thermalization from perturbative approaches\cite{Baier2001}, and there is a slight delay in the onset of thermalization. Now such model has been generalized to the bulk geometry with electrostatic potential~\cite{GS,CK,Yang1}, high curvature corrections~\cite{Zeng2013,Zeng2014,Baron,Li3764}, and some other gravity models~\cite{Baron1212,Arefeva,Hubeny,Arefeva6041,Balasubramanianeyal6066,Balasubramanian4,Balasubramanian3,Balasubramanian9,Fonda,Cardoso,Hubeny2014,Zeng2015,Giordano,Zhang1,Camilo,Giordano1,Zeng20152}.

In this paper, we intend to study holographic thermalization in the bulk surrounded by phantom dark energy. The phantom dark energy is the scalar field with a negative kinetic term with an equation of state parameter $\omega<-1$. As already known, the presence of cosmological phantom fields continues to receive supports from both collected
observational data~\cite{Koma} and theoretical models~\cite{dyn}. All these programs have pointed out an accelerated expansion of the universe, dominated by an exotic fluid of negative pressure. Furthermore, there are
evidences suggesting the exotic fluid could be of phantom nature~\cite{nat1,nat2}. Since then, an interest in phantom fields has grown and resulted in many phantom black hole solutions~\cite{gr,gerard2}. In recent years, many issues pertaining to phantom black hole, such as thermodynamic stability~\cite{thermo} and light paths~\cite{light,light2} have been dealt with. In this paper, we study holographic thermalization in the spacetime dominated by phantom dark energy. The authors in ref.\cite{Zeng20152} has studied the holographic thermalization dominated by quintessence dark energy and their results show that the smaller the state parameter
of quintessence is, the harder the plasma to thermalize. Considering the first year WMAP data together with the 2dF galaxy survey and the supernova Ia data favor the phantom energy over the cosmological constant and the quintessence, here we study the effect of phantom dark energy on the holographic thermalization. Our results show that the sate parameters of dark energy play different roles in the process of holographic thermalization. Our results here shows that the smaller the phantom dark energy parameter is, the easier the plasma to thermalize when the chemical potentieal is fixed, which is different from the case produced by quintessence dark energy.

The structure of this paper is outlined as follows. In Section~\ref{Nonlocal_observables}, we will briefly review the gravitational collapse solution in the spacetime dominated by the phantom dark energy. In Section\ref{Numerical_results}, we derive the equations of motion of the minimal area surface theoretically firstly, and then perform a systematic analysis about how the dark energy parameters and chemical potential affect the thermalization time resorting to numerical calculation. We also study the thermalization velocity and thermalization acceleration with the fitting functions of the thermalization probes. The last section is devoted to discussions and conclusions.


\section{Gravitational collapse surrounded by phantom dark energy}\label{Nonlocal_observables}
Start from the action
\begin{eqnarray}
 S=\int d^4x\sqrt{-g}[R+2\eta F^{\mu\nu}F_{\mu\nu}+2\Lambda],\label{action}
\end{eqnarray}
reference \cite{Jardim}  recently obtained a phantom Reissner-Nordstrom AdS black hole solution
\begin{equation}
 ds^{2}=-f(r)dt^{2}+f^{-1}(r)dr^{2}+r^{2}d \phi ^{2}+r^{2}\sin^{2}\phi d\varphi^{2},\label{metric}
\end{equation}
where%
\begin{equation}
 f(r)=1-\frac{2M}{r}+\eta \frac{q^2}{r^2}+\frac{r^2}{L^2}.  \label{areacoorections2}
\end{equation}%
In which $L$ is the radius of the AdS black hole, $\eta$ is the phantom dark energy parameter. For $\eta=1$, the solution in
Eq.(\ref{metric}) represents for the Reissner-Nordstrom-AdS black hole.
There is also a purely electric gauge field given by
\begin{equation}
A_t=(u-\frac{q}{r}), \label{au}
\end{equation}%
where $u$ is a constant corresponding to the electrostatic potential at $r\rightarrow\infty$, which will
be related to the chemical potential in the dual gauge theory according to the AdS/CFT
correspondence. It is defined such that the gauge field vanishes at the horizon, namely $u=q/r_h$.

In the
context of the AdS/CFT correspondence, one is often interested in dual gauge theories living
on flat space. In order to achieve this goal, one often use the infinite volume limit\cite{Chamblin}. After this step,  we can get the phantom Reissner-Nordstrom-AdS black brane solution
\begin{equation}
ds^{2}=-F(r)dt^{2}+F^{-1}(r)dr^{2}+r^{2}dx_i^{2},\label{bh}
\end{equation}
in which  $i=1,2$, and
\begin{equation}\label{blackening}
F(r)=-\frac{2M}{r}+\eta \frac{q^2}{r^2}+\frac{r^2}{L^2}.
\end{equation}
In this case, the event horizon, defined by $F(r_h)=0$,  is plane
instead of sphere. According to the definition of the surface gravity, we also can get the Hawking temperature
emitting from the black brane
\begin{equation}\label{blackening}
T=\frac{\kappa}{2\pi}=\frac{1}{4\pi}(\frac{2M}{r^2_h}-\eta \frac{2q^2}{r_h^3}+\frac{2r_h}{L^2}),
\end{equation}
which in the framework of AdS/CFT can be viewed as
the equilibrium temperature of the dual field theory living on the boundary.

To get a Vaidya type evolving black brane, we will reexpress the metric in Eq.(\ref{bh})
in Eddington-Finkelstein coordinate $dv=dt+\frac{dr}{F(r)}$,
and then rewrite it with the  inverse radial coordinate $z=\frac{L^2}{r}$.
The background spacetime in Eq.(\ref{bh}) changes into
\begin{equation}
ds^2=\frac{1}{z^2} \left[ - H(z) d{v}^2 - 2 dz\ dv +
dx_i^2 \right], \label{collpse}
\end{equation}
where $L$ has been set to $1$, and
\begin{equation}
H(z)=1-2 M z^3+\eta q^2 z^4.\label{h}
\end{equation}
Treating the mass parameter in Eq.(\ref{h}) as an arbitrary function of $v$, Eq.(\ref{collpse}) can be
regarded as a gravitational collapse solution surrounded by phantom dark energy\cite{Balasubramanian1,Balasubramanian2,GS,CK}. This solution
is not a solution of the action  in Eq.(\ref{action}) anymore. It includes the contributions of some
external matter fields.
 As one can show,
such a metric is sourced by the null dust with the energy momentum tensor flux and gauge flux \cite{GS,CK}
\begin{eqnarray}
  &&8\pi G T_{\mu\nu}^{matter}=z^{2}[\dot{M}(v)-2zq(v)\dot{q}(v)]\delta_{\mu v}\delta_{\nu v},\nonumber\\
  &&8\pi G J^{\mu}_{matter}=z^4 \dot{q}(v)\delta^{\mu z},
\end{eqnarray}
where the dot stands for derivative with respect to coordinate $v$,  $M(v)$ and  $q(v)$ are
 the mass and charge of a collapsing black brane


\section{Holographic thermalization} \label{Numerical_results}

In the previous section, we have got a gravitational collapse solution which describes the collapse of a thin-shell of charged dust from the boundary toward
the bulk interior of asymptotically anti-de Sitter spaces. According to the AdS/CFT correspondence, this process is dual to the
thermalization of plasma of quarks and gluons which are formed  in heavy ion colliders such as the RHIC
and LHC. To describe the thermalization more comprehensive, we should have an initial state and an equilibrium state, which is dual to a pure AdS  and a stationary black brane respectively. Recent investigations show that as the
mass $M(v)$ and charge $q(v)$ are written as the smooth functions \cite{Balasubramanian1,Balasubramanian2,GS,CK}
\begin{equation}
M(v) = \frac{M}{2} \left( 1 + \tanh \frac{v}{v_0} \right),
\end{equation}
\begin{equation}
q(v) = \frac{q}{2} \left( 1 + \tanh \frac{v}{v_0} \right),
\end{equation}
where $v_0$ represents a finite shell thickness, one can construct such a gravity model. In this case,
in the limit $v\rightarrow
-\infty$, the mass vanishes and the background  in Eq.(\ref{collpse}) corresponds to a pure AdS space. While in
the limit $v\rightarrow \infty$, the mass is a constant and the
 background corresponds to a phantom Reissner-Nordstrom-AdS black brane. Thus for different values of the time $v$, the
background  in Eq.(\ref{collpse}) stands for different stages of gravitational collapse, which represents for different stages
of thermalization in the dual conformal field theory.

\subsection{Nonlocal probes}
As the gravity model describing the thermalization is constructed, we will choose proper observables to probe it.
Since
local observables in the boundary such as expectation values of the energy-momentum tensor
are not sensitive to the thermalization process, one needs to consider the non-local
observables such as the two-point correlation functions, expectation values of  Wilson loops,
and entanglement entropy. Recent investigations show that all of them have similar behavior, thus we are interested only in the
 expectation values of  Wilson loops. Wilson loop operator is defined as a path ordered integral of gauge field over a closed contour,
 and its expectation value is approximated geometrically  by the AdS/CFT correspondence as \cite{Balasubramanian2,Maldacena80}
\begin{equation} \label{area}
\langle W(C)\rangle \approx e^{-\frac{A_{ren}(\Sigma)}{2\pi\alpha'}},
\end{equation}
where $C$ is the closed contour, $\Sigma$ is the minimal bulk surface ending on $C$ with $A_{ren}$ its renormalized
minimal  area surface, and  $\alpha'$ is the Regge slope parameter.

Here we are focusing solely on the rectangular space-like Wilson loop. In this case, the enclosed rectangle can  always be
chosen to be centered at the coordinate origin and lying on the $x_1-x_2$ plane with the assumption that the corresponding
 bulk surface is invariant along the $x_2$ direction. The area of the minimal  area surface is written usually as \cite{Wald}
\begin{equation}
A=\int \sqrt{g_{\mu\nu}dx^{\mu}dx^{\nu}}dx\int dy,
\end{equation}
 here for notational simplicity, $x_1$ has been
replaced by $x$ and $x_2$ has been
replaced by $y$.
 For the  Vaidya-like AdS black branes in  Eq.(\ref{collpse}), the area of the minimal  area surface can be written as
\begin{equation}\label{lequation}
A=\int_{\frac{l}{2}}^{\frac{l}{2}}dx
\frac{\sqrt{1-2z'(x)v'(x) - H(v,z) v'(x)^2}}{z(x)^2},
\end{equation}
where $l$ is the boundary septation along  $x$ direction.
As to Eq.(\ref{lequation}), we can treat the integral as a   Lagrangian $\cal{L}$ and with it we can get the equations of motion of $z(x)$  and  $v(x)$
\begin{eqnarray} \label{aequation}
0&=&4-4v'(x)^2H(v,z)-8v'(x)z'(x)-2z(x)v''(x)\nonumber\\
&+&z(x)v'(x)^2\partial_zH(v,z),\\
0&=&v'(x)z'(x)\partial_zH(v,z)+\frac{1}{2}v'(x)^2\partial_vH(v,z)\nonumber\\
&+&v''(x)H(v,z)+z''(x).
\end{eqnarray}
To solve these equations, we need
to consider the symmetry of the minimal  area surface and impose the following boundary conditions
\begin{equation}\label{initial}
z(0)=z_{\star},  v(0)=v_{\star} , v'(0) =
z'(0) = 0.
\end{equation}
In addition,  the
area $A$ is divergent due to its contribution near the AdS boundary. To eliminate it, we should impose a cutoff near the boundary \begin{equation}\label{regularization}
z(\frac{l}{2})=z_0, v(\frac{l}{2})=t_0,
\end{equation}
where $z_0$ is the IR radial cut-off and $t_0$ is the time that the  minimal area surface approaches to the boundary, which is called as the thermalization time usually.
AS the divergent part of $A$ is subtracted, the  minimal  area surface, called
renormalized minimal  area surface, can be cast into
\begin{equation}\label{aren}
A_{ren}=2\int_0^{\frac{l}{2}}dx
\frac{z^2_{\star}}{z(x)^4}-\frac{2}{z_0}.
\end{equation}
Note that here we have used the conserved equations obtained from the Lagrangian $\cal{L}$ in  Eq.(\ref{lequation}), and $\frac{2}{z_0}$ is the contribution of the minimal  area surface in the AdS boundary\cite{Balasubramanian1,Balasubramanian2}.

\subsection{Numerical results}

In this section,  we will solve the equations of motion of
minimal  area surface  numerically, and then explore how the chemical potential and  phantom dark energy coefficient affect
 the thermalization.  During the numerics, we will take the shell thickness and  UV
 cut-off as $v_0 = 0.01$, $z_0 = 0.01$ respectively.

\begin{figure}
\centering
\subfigure[$$]{
\includegraphics[scale=0.75]{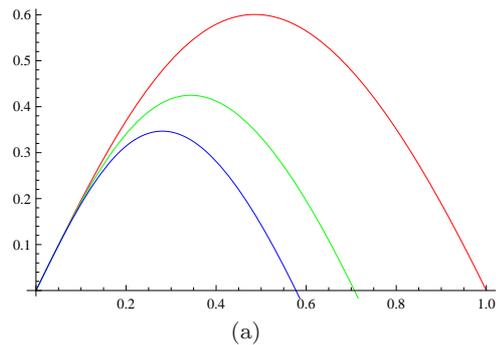}}
 \caption{\small Change of $u/T$ to the charge $q$. The red line, green line and blue line correspond to $q=-1,-2,-3$ respectively.} \label{fig0}
\end{figure}

 According to the AdS/CFT correspondence, we know that the electromagnetic
field in the bulk is dual to the chemical potential in the dual quantum field theory, therefore we will use the  electromagnetic
field defined in Eq.(\ref{au}) to explore the effect of the chemical potential on the thermalization process in the AdS boundary.
 When we investigate the effect of  the chemical potential
 on the thermalization, the quantity
which is physically meaningful is the ratio of
$u/T$, where $T$ is the Hawking temperature of the phantom Reissner-Nordstrom-AdS black brane.
 However, as stressed in \cite{GS,Ling2013},
 the chemical potential has energy
units in the dual field theory $([u]= 1/[L])$ while $A_{\mu}$
defined in Eq.(\ref{au}) is dimensionless. Thus one has to redefine the  electromagnetic
field as
$\tilde{A}_{\mu}=A_{\mu}/p$, where $p$
is a scale with length unit that depends on the particular compactification.
Therefore, through this paper we will use the following ratio
\begin{equation} \label{ratio}
\frac{u}{T}=\frac{\lim\limits_{r\rightarrow\infty} \tilde{A}_{\mu}}{T}=\frac{2q (1+\eta q^2)}{p (3-\eta q^2)},
\end{equation}
to check the effect of the chemical potential on the thermalization time, in which we have set the horizon
$r_h=1$. For simplicity, we will call the ratio in Eq.(\ref{ratio}) as chemical potential hereafter.
For the phantom dark energy parameter, we know that it satisfies the condition $\eta \leq -1$, in
 this paper, we will choose $\eta=-1,-2, -3$ in our numerical result. In addition, we know that when
 $\eta=1$, the solution in  Eq.(\ref{metric})  stands for the Reissner-Nordstrom-AdS black brane. To compare conveniently, we
 will also study the case  $\eta=1$.
From  Eq.(\ref{ratio}), we know that for $\eta=1$, the chemical potential raises from $0\rightarrow \infty$
 provided $q $ changes from 0 to $\sqrt{3}$. While for $\eta \leq -1$, the chemical potential is not monotonous as the charge
 increases, please see Figure (\ref{fig0}). For  $\eta=-1,-2, -3$, the chemical potential increases monotonously for $0<q<0.49, 0<q<0.35, 0<q<0.28$ respectively, and monotone decreasing for the other values. To investigate conveniently, we will choose $q=0.5, 0.7,0.9$ in this paper. With this choice, from figure (\ref{fig0}) we know that  for the negative phantom dark energy parameters, as the charge
increases, the chemical potential decreases, which is different from the case $\eta=1$ where the chemical potential increases when the charge increases.

With the initial conditions in Eq.(\ref{initial}), we will solve  $z(x)$  and  $v(x)$ firstly.
When we are interested in the
effect of chemical potential on the thermalization, the phantom dark energy coefficient is fixed, and when we are interested in the
effect of   phantom dark energy  coefficient on the thermalization, the  chemical potential is fixed.
Since different initial time $v_{\star}$
corresponds to different stages of the thermalization, we will study whether the phantom dark energy coefficient
and chemical potential have the same effect on the thermalization for different $v_{\star}$.

\begin{table}
\begin{center}\begin{tabular}{l|c|c|c|c}
 \hline
\multirow{4}{1cm}{$v_{\star}=-0.888$}
          &            &  $q=0.5$       &  $q=0.7$        &    $q=0.9$     \\
  &$\eta=1$           & 0.989522     & 0.999862       & 1.01398     \\
   &$\eta=-1$           &  0.969131     & 0.9597     & 0.947474   \\
     &$\eta=-2$          & 0.959202     &  0.94115      &0.918376      \\
   &  $\eta=-3$           &0.949734      &  0.923542    & 0.891606     \\ \hline
   \multirow{4}{1cm}{$v_{\star}=-0.555$}
  &$\eta=1$        &  1.34905      & 1.37703      &  1.41573     \\
   &$\eta=-1$           &  1.29452    & 1.26975      & 1.23855     \\
     &$\eta=-2$        & 1.26889     & 1.22264       & 1.16637      \\
   &  $\eta=-3$           & 1.24432      & 1.17899       &   1.10331     \\ \hline
 \multirow{4}{1cm}{$v_{\star}=-0.333$}
  &$\eta=1$        & 1.52816     &  1.56579       &  1.61914     \\
   &$\eta=-1$         & 1.45507    & 1.42259      & 1.38149     \\
     &$\eta=-2$         & 1.42132      &1.36125      & 1.29001     \\
   &  $\eta=-3$            &  1.38917      &  1.30584       & 1.21261     \\ \hline
\end{tabular}
\end{center}
\caption{The thermalization time $t_0$ of the  minimal area surface  probe for different phantom dark energy
 coefficient  and different charge at $v_{\star}=-0.888,-0.555, -0.333$ respectively.}\label{tab1}
\end{table}
The result is listed in Table (\ref{tab1}), from it, we can observe the following phenomena:
\begin{description}
\item$\bullet$ For a fixed  chemical potential, as the phantom dark energy  parameter decreases, the thermalization time increases, which means that the small  dark energy  parameters delay the thermalization;
\item$\bullet$ As the charge increases, the thermalization time increases for the positive
 dark energy  parameter while  it decreases for the negative parameters. Namely  the charge has oppositive effect on the thermalization time for the positive and negative parameters. However, as the charge raises, we know that  the chemical potential increases for the positive parameter, and decreases for the negative parameters. So we conclude that for all the phantom dark energy  parameter, as the chemical potential raises, the thermalization time increases.
\item$\bullet$ For different initial thermalization time, the effect of the phantom dark energy  parameter and the chemical potential on the thermalization are the same. Thus we can predict that this effect would be the same to the whole thermalization time.
\end{description}

\begin{figure}
\centering
\subfigure[$q=0.5$]{
\includegraphics[scale=0.75]{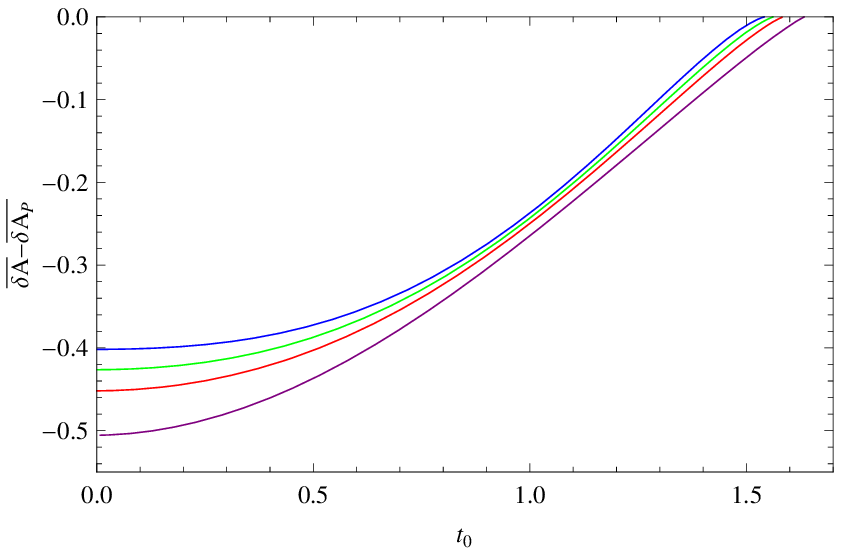}
}
\subfigure[$q=0.7$]{
\includegraphics[scale=0.75]{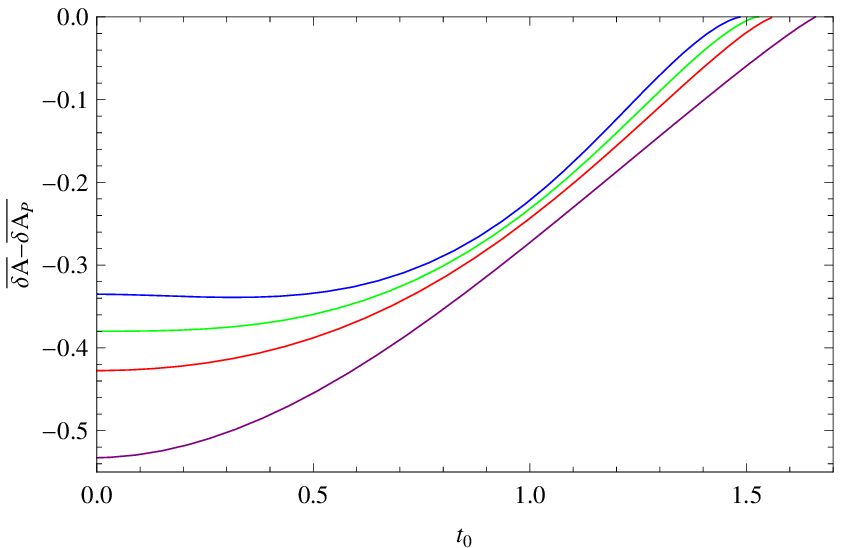}  }
\subfigure[$q=0.9$]{
\includegraphics[scale=0.75]{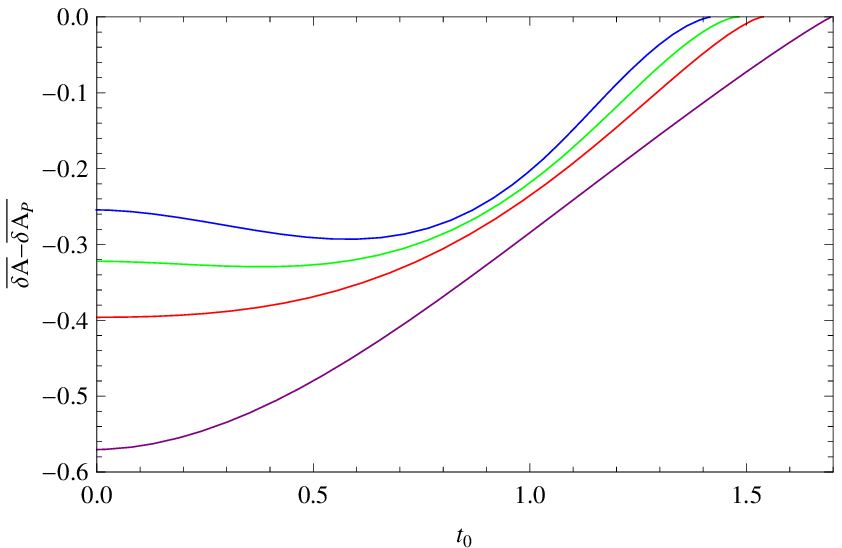}  }
 \caption{\small Thermalization of the renormalized minimal area surface   for different phantom dark energy parameters  with the same chemical potential at the same boundary separation $l=2.2$.
 The purple line, red line, green line,  and blue line correspond to  $\eta=1, -1, -2, -3$ respectively.} \label{fig1}
\end{figure}


\begin{figure}
\centering
\subfigure[$\eta=1$]{
\includegraphics[scale=0.75]{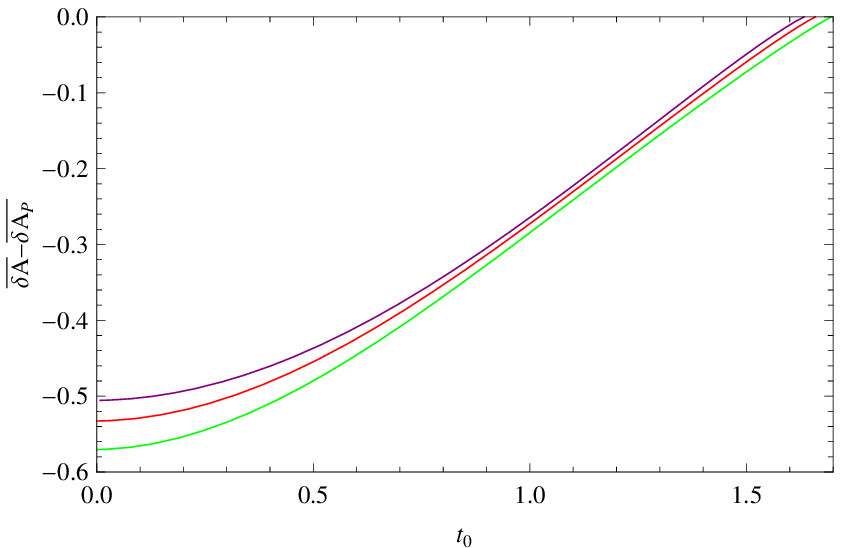}  }
\subfigure[$\eta=-1$]{
\includegraphics[scale=0.75]{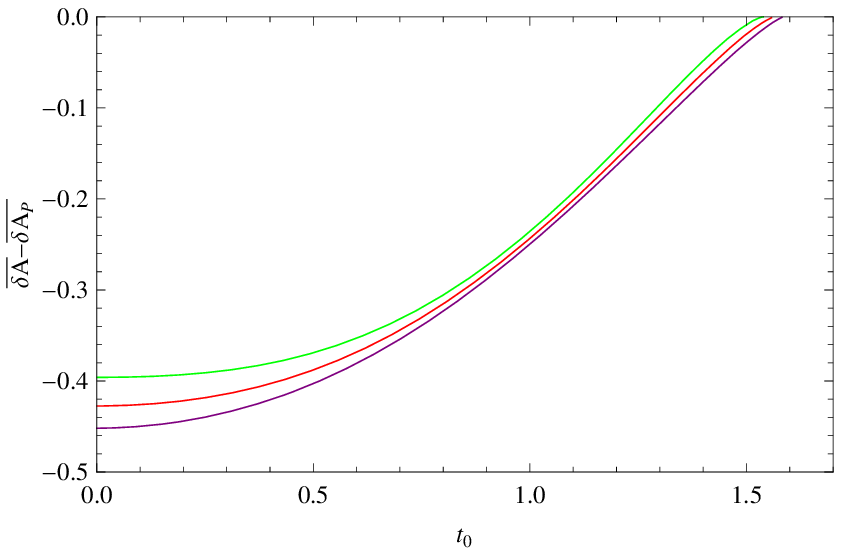}
}
\subfigure[$\eta=-2$]{
\includegraphics[scale=0.75]{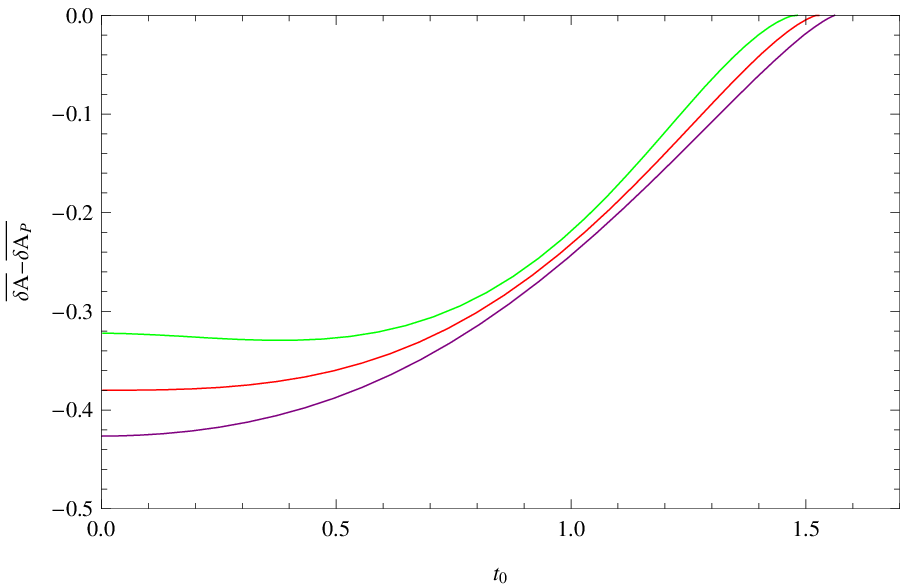}  }
\subfigure[$\eta=-3$]{
\includegraphics[scale=0.75]{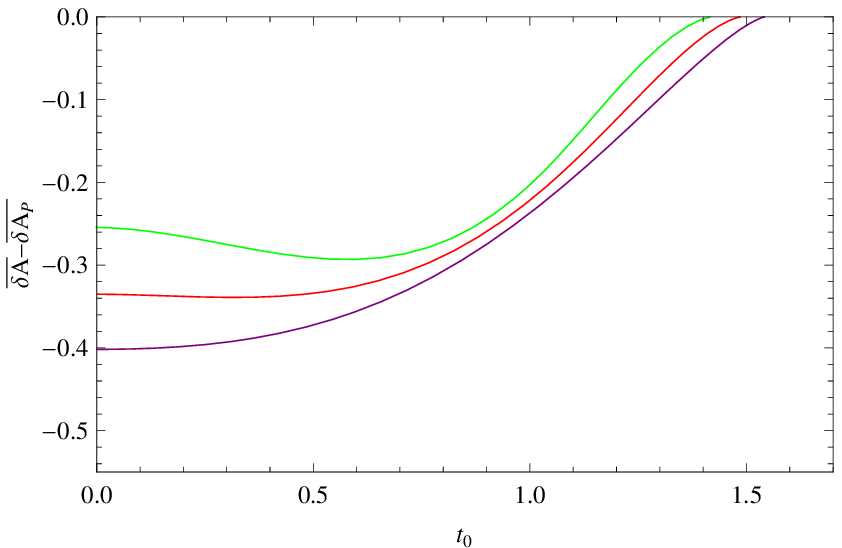}  }
 \caption{\small Thermalization of the renormalized minimal area surface   for different  chemical potentials  with the same  phantom dark energy parameter at the same boundary separation $l=2.2$.
 The purple line, red line, and green line correspond to  $0.5,0.7,0.9$ respectively.} \label{fig2}
\end{figure}

\begin{figure}
\centering
\subfigure[$\texttt{}$]{
\includegraphics[scale=0.75]{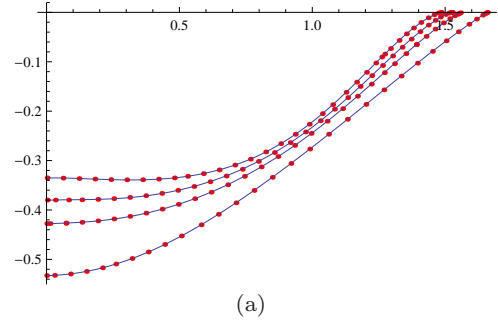} }
 \caption{\small Comparison of the thermalization process between the numerical curves and fitting functions.} \label{fig3}
\end{figure}

\begin{figure}
\centering
\subfigure[$\texttt{}$]{
\includegraphics[scale=0.75]{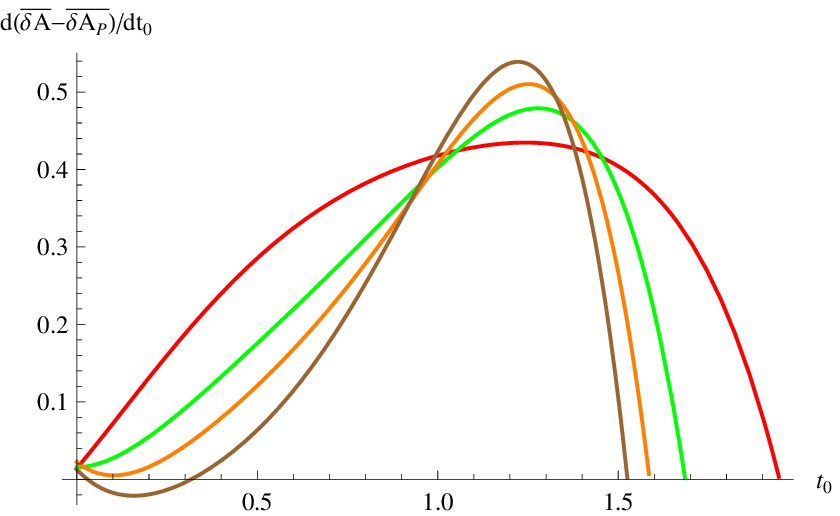} }
 \caption{\small Thermalization velocity of the renormalized   minimal area surface at $q=0.7$. The red line, green line,  orange line, and brown line  correspond to  $\eta=1, -1, -2, -3$ respectively.} \label{fig4}
\end{figure}

\begin{figure}
\centering
\subfigure[$\texttt{}$]{
\includegraphics[scale=0.75]{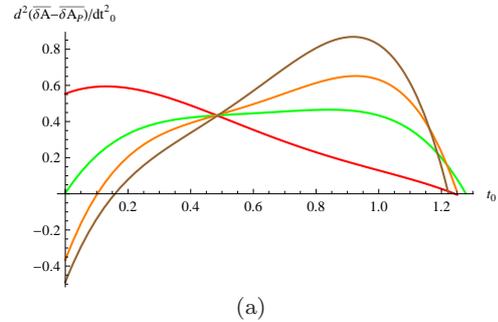} }
 \caption{\small Thermalization acceleration of the renormalized   minimal area surface at $q=0.7$. The red line, green line,  orange line, and brown line  correspond to  $\eta=1, -1, -2, -3$ respectively.} \label{fig5}
\end{figure}
With the numeric result of $z(x)$, we can further get the renormalized minimal  area surface defined in   Eq.(\ref{aren}). The result is shown in Figure (\ref{fig1}) and Figure (\ref{fig2}). In  Figure (\ref{fig1}), we fix the chemical potential to investigate the effect of the phantom dark energy parameter on the thermalization, while in  Figure (\ref{fig2})  we fix the  phantom dark energy parameter  to investigate the effect of  chemical potential on the thermalization. We are interested in the $l$ independent quantity $\overline{\delta A}\equiv A_{ren}/l$,
 $\overline{\delta A_{q}}\equiv A_{q}/l$, where $A_{q}$ is the renormalized minimal area surface at the equilibrium state. To observe the thermalization conveniently, we plot the quantity
$\overline{\delta A}-\overline{\delta A_{q}}$. In this case, the thermalized state is labeled by the null point of $\overline{\delta A}-\overline{\delta A_{q}}$. In each picture, the vertical axis indicates the renormalized minimal area surface while the horizontal axis indicates the thermalization time $t_0$.

From Figure (\ref{fig1}), we know that as the phantom dark energy parameter decreases, the thermalization time  increases. That is, the small  phantom dark energy parameter  promotes the thermalization. This phenomenon is more obvious as the charge raises.
Especially for the case $q=0.9$, we find the difference of the thermalization curve for different phantom dark energy parameters is most obvious.
The effect of the chemical potential  on the thermalization is plotted in Figure (\ref{fig2}). For (a) in Figure (\ref{fig2}),
the background recovers to the Vaidya AdS black brane with a chemical potential which was investigated in \cite{GS,CK,Yang1}. We see that as the chemical potential  increases, the thermalization time enhances. That is, the chemical potential delays the thermalization, which is consistent with that obtained in \cite{GS,CK,Yang1}. For (b),(c),(d) in Figure (\ref{fig2}), we find that   as the charge  increases, the thermalization time decreases. This effect is not the same as that in (a).  Note that for the case $\eta=-1,-2,-3$, the chemical potential is  monotonously decreasing as the charge increases. Thus we also can conclude that as the chemical potential  increases, the thermalization time raises too. Namely the chemical delays the thermalization always though this seems different for the negative parameters and positive parameters as the charge increases.

In addition, we find the numeric curves for the thermalization process can be fitted as  analytical functions with respect to the thermalization time. Thereafter, we mainly focus on the effect of the phantom dark energy parameter on the  thermalization.  As an example, we consider the case $q=0.7$. The fitting functions of the thermalization curves for $\eta=1, -1, -2, -3$ can be expressed as
\begin{eqnarray}
\begin{cases}
 -0.532952+0.0142505 t_0+0.276771 t_0^2+0.111791 t_0^3\\
  ~~ ~~ ~~ ~~~~~~~~-0.260812 t_0^4+0.151048 t_0^5-0.0328631 t_0^6\\
-0.427766+0.0156803 t_0+0.00149207 t_0^2+0.442568 t_0^3\\
  ~~ ~~ ~~ ~~~~~~~~-0.52845 t_0^4+0.346392 t_0^5-0.0937172 t_0^6\\
-0.380358+0.0222836 t_0-0.18342 t_0^2+0.76406 t_0^3\\
  ~~ ~~ ~~ ~~~~~~~~-0.913827 t_0^4+0.638737 t_0^5-0.179835 t_0^6\\
 -0.335312+0.0126767 t_0-0.244621 t_0^2+0.73413 t_0^3\\
  ~~ ~~ ~~ ~~~~~~~~ -0.859683 t_0^4+0.690742 t_0^5-0.219645 t_0^6
\end{cases}
\end{eqnarray}
 Figure (\ref{fig3}) is the comparison of the numerical curves and fitting function curves. It is obvious that at
the order of $t_0^6$, the thermalization curves can be described well by the fitting functions.
 With the analytical functions
of the thermalization curves, we can get the thermalization velocity,  and thermalization acceleration conveniently.
From the thermalization velocity curves in Figure (\ref{fig4}), we find the thermalization process is similar for different phantom dark energy parameters, namely at the middle stage of the thermalization process
there is a phase transition point which divides the thermalization into an acceleration phase and a deceleration phase. This result is reasonable because the thermalization can not accelerate always in order to approach to an equilibrium state.
The phase transition points can also be read off from the null point of the
acceleration curves, which are plotted in Figure (\ref{fig5}). For the negative phantom dark energy parameter, we find as the parameter decreases, the phase transit decreases.
 For all the phantom  dark energy  parameters, we find the initial acceleration of renormalized minimal area surface decreases as the state parameter decreases. Correspondingly, the initial thermalization velocity of minimal area surface decreases gradually. In addition, from  Figure (\ref{fig4}) and  Figure (\ref{fig5}) we find the maximum acceleration and maximum velocity increase as the phantom  dark energy  parameters decrease.

\section{Conclusions} \label{sec:conclusions}
We construct a gravitational collapse model which is interpolated  between
a pure AdS space and a phantom Reissner-Nordstrom AdS black brane. We probe this collapse behavior with the minimal area surface.
According to the language of AdS/CFT correspondence,  this is dual to probe the thermalization behavior of the quark gluon plasma by the expectation value of Wilson loop in the dual conformal field theory. We  focus mainly on the effect of phantom dark energy and chemical potential on the thermalization time. We find for a fixed chemical potential,  the smaller the phantom dark energy parameter is, the easier the plasma thermalizes. And for a fixed dark energy parameter, the larger the  chemical potential is, the harder the plasma to thermalize. For the effect of chemical potential on the thermalization time, there are two points should be stressed:
\begin{description}
\item$\bullet$ From  Figure (\ref{fig2}), we know that as the charge raises, the thermalization time enhances for the case $\eta=1$, while it decreases for the case $\eta=-1,-2,-3$. However, in the dual conformal field theory, the quantity which is meaningful physically is the chemical potential not the charge.
    With the consideration that for $q>0.5$, the chemical potential raises for $\eta=1$, while it decreases for $\eta=-1,-2,-3$ as the charge grows. Thus \emph{in the parameter range} which we choose, we can conclude that the chemical potential delays the thermalization for all the phantom dark energy parameters.
\item$\bullet$ From  Figure (\ref{fig0}), we know that the chemical potential is not a  monotone function with respect to the charge. For $0<q<0.49, 0<q<0.35, 0<q<0.28$, the chemical potential  is monotonously increasing as the charge raises for
      $\eta=-1,-2, -3$ respectively. Thus, if we choose the smaller charge, e.g. $q=0.2$, we find the chemical potential promotes the thermalization, which is different from the previous result that the chemical potential delays the thermalization. So strictly speaking, the effect of the chemical potential on the thermalization time depends on the dark energy parameters.
\end{description}

Besides our previous investigation, we also obtain the fitting functions of the thermalization curves, and with the functions, we further investigate the thermalization velocity and thermalization acceleration. From the thermalization velocity curves, we know that at the middle stage of the thermalization there is a phase transition point which separates the thermalization into an acceleration phase and a deceleration phase. The phase transition point is found to be shifted for different dark energy parameters.
Especially for  $\eta=-1,-2,-3, $ we find he phase transit decreases as the phantom dark energy parameter decreases. Both the maximum acceleration and maximum velocity are found to be increased as the dark energy  parameters decrease.
Though our study in this part is not so strict as that in \cite{liu1,liu2} , where the whole thermalization was divided
into four processes and the analytical function for each process was given strictly, our investigation can provide another sight
to understand the thermalization process.

\begin{acknowledgements}
  This work is supported  by the National
 Natural Science Foundation of China (Grant No. 11405016 and No. 11205226).

\end{acknowledgements}



\end{document}